\documentclass[12pt]{article}
\usepackage{amsmath,amsfonts,amssymb,graphicx,epsfig}
\usepackage[latin1]{inputenc}
\usepackage{colordvi}
\usepackage[dvips]{color}
\def\ba{\begin{eqnarray}}

\def\ea{\end{eqnarray}}

\def\nn{\nonumber}

\begin{document}
\title{\bf Magnetic vortex-like excitations on a sphere}
\author{G.S. Milagre and Winder A. Moura-Melo\thanks{E-mails: winder@ufv.br, winder@pq.cnpq.br}\\
\small \it Departamento de F\'{\i}sica, Universidade Federal
de Vi\c{c}osa\\ \small \it 36570-000, Vi\c{c}osa, Minas
Gerais, Brazil}

\date{}
\maketitle
\begin{abstract}
We study magnetic vortex-like solutions lying on the spherical surface. The simplest cylindrically symmetric vortex presents two cores (instead of one, like in
open surfaces) with same charge, so repealing each other. However, the net vorticity is computed to vanish in accordance with Gauss theorem. We also address the problem of a flat plane in which a conical, a pseudospherical and a
hemispherical segments were incorporated. In this case, if we allow the vortex to move without appreciable
deformation in this support, then it is attracted by the conical apex and by the pseudosphere as well, while it is repealed by the hemisphere. This suggests that such surfaces could be viewed as pinning and depinning geometries for those excitations.
Spherical harmonics coreless solutions are discussed within some
details.
\end{abstract}
\newpage
\section{Introduction and Motivation}
Spheric-like shapes abound Nature: From micro to macroscale, particles and bodies with this geometry fill the Universe. Besides, spherical symmetry is very important in Physics and other sciences and is behind a number of triumphs of theoretical physics, namely the analytical resolution of the Hydrogen atom, the Schwarzschild and Kerr solutions of the Einstein field equations of General Relativity, and so forth.\\

Recent advances in Nanoscience and Nanotechnology have led, in turn, to several structures at
extremely small scales displaying a sort of shapes: cylindrical
and toroidal nanotubes, nanocones, and spherical nanoshells are some examples.
Namely, the latter has received a great deal of efforts in the last years.
Actually, a novel paradigm in bottom-up design for fabricating new materials
lies on the geometry and topology of the sphere \cite{sphere-new-methods}, the
study of optical and electronic properties of metallic nanoshells has revealed a
number of results which could be important for spectroscopy based upon surface
plasmon resonance\cite{Prodan}, and spherical cavities may provide a smooth
mechanism for pinning superconducting line vortices
\cite{sphere-superc-vortex}. Concerning topological excitations
in this framework, we may quote the formation and dynamics of vortex-like
magnetisation observed in chains of spheres \cite{vortex-chain-sphere} and in
hollowed spherical objects \cite{sphere-cr-sizes}. The relevance of topological objects lying on the spherical geometry has also been considered in nematic shells textures, where a transition from half-disclinations to half-hedgehogs excitations is predicted as the thickness of the shells increases\cite{Vitelli-Nelson}.\\

Besides these examples, excitations with non-trivial topology have a long history in Physics
and are related to important phenomena. A classical example is the
(topological) phase transition provided by vortex-pair dissociation in a
number of planar-like samples, like magnets and superfluids \cite{BKT}. More
recently, the manipulation of curling magnetisation in bi-hollowed magnetic
nanodots has been investigated for producing basic logical operations by means
of the switching of the vortex between the two cavities
(antidots)\cite{nanodot}. Other pseudoparticles, generically called solitons,
appear as classical and topologically stable solutions carrying finite size
and mass in a number of non-linear models. For example, two-dimensional (2D)
isotropic Heisenberg ferromagnets exhibit, in their continuum limit, spin
textures of the solitonic type\cite{BP}. Such excitations are also present in
theories of more fundamental interactions, like the skyrmions participating in
the effective strong coupling \cite{skyrmion} and exotic particles carrying
magnetic charge\cite{tHP}.\\

Solitons, vortices and other topological excitations, particularly associated with 2D magnetic systems, are often
investigated in the lattice and continuum frameworks of Heisenberg-like
models. Although there is a extensive literature dealing with such a subject in the planar case, the same cannot be said about more complex geometries, say, curved, torsed, and so on. Motivated, in part, by the fact that the fabrication and manipulation of magnetic surfaces with several shapes are currently facilitated, a number of works has been
devoted to such issues. For example, solitons have been considered in spins
systems displaying the shapes of
cylinders\cite{soliton-cylinder,soliton-cylinder-sphere},
spheres\cite{soliton-cylinder-sphere}, and cones \cite{soliton-cone}. Magnetic vortices of the XY model have been studied on the surface
of a cone\cite{vortex-cone}, in which one realizes that the vortex is attracted towards the apex. On the other hand, on the pseudosphere (a hyperbolic surface with constant Gaussian curvature) soliton appear to be unstable, decaying against the ground-state, while vortex-like solutions present the interesting feature of having finite energy at the thermodynamical limit\cite{pseudonosso}. The main
lesson left by these works is that such excitations present some characteristics that strongly depend on the geometrical parameters associated to the supports. \\

Here, we shall be interested in analysing the basic properties of magnetic
vortices whenever they lie on a spherical surface. It should be stressed that
some analogues of this system are very important and have received much
attention, namely in Physics of Fluids. For instance, a number of features accompanying the structure and
dynamics of actual hydrodynamic curling fields (hurricanes, etc) in the Earth
atmosphere are somewhat obtained from the study of vortex-like solutions on a
sphere\cite{Crowdy}. [Similar plans are claimed to work for other planetary
``atmospheres'']. However, it should be emphasised that the latter are
different from those associated to the spin field in the sense that these are
not vectors in the physical space, so they do not experience the curvature
effects directly, say, by means of the connection (Christoffell symbols) in
the covariant derivative. Indeed, although this coupling is nonvanishing it
does not contribute to the magnetic energy of those excitations: It is an
extra energy called magnetoelastic\cite{SaxenaPhysA} which is important for
studying deformable surfaces like those appearing in soft condensed matter
problems.\\

In order to carry out our investigation we shall employ the continuum limit of
the Planar Rotator Model ($O(2)$ isotropic model, or alternatively, the XY regime with spins confined to the internal plane). Among other, we realize the appearance of two singular cores for the simplest solution, so that the net vorticity on the
whole sphere vanishes, in contrast to what is observed in the plane (or in other non-compact surfaces). Indeed, whenever measured isolated, the cores present the same charge, in such a way that they repeal each other. The energy is computed and may be written in a form which resembles that of a
vortex in the plane, provided that the sphere radius is large enough. In addition, we consider a very slow (adiabatic) dynamic for a vortex on a large plane in which were inserted a hemisphere, a truncated cone and a part of a pseudosphere. Now, our intention is to analyse how these inserted
geometries affect vortex motion, say, attracting or repelling it in
comparison with the flat surface. We have seen that if the interfaces between
these geometries are smooth and the in-plane vortex does not change its
profile (no large deformation neither it develops out-of-surface component)
then the hemisphere repeal the vortex, while in the other cases it is attracted. Attention is also paid to other solutions which present no singular cores, the well-known spherical harmonics, whose energies are analytically evaluated and are finite. Finally, we put forward our Conclusions and Prospects for future investigation. 

\section{The continuum Heisenberg model on the sphere}

The anisotropic Heisenberg model, for nearest-neighbour interacting spins, on a
two-dimensional lattice, is given by the Hamiltonian below:
 \ba H_{\rm latt}=-J'\,\sum_{<i,j>} {\cal
H}_{i\,,j}=-J'\,\sum_{<i,j>}
(S^x_iS^x_j+S^y_iS^y_j+(1+\lambda) \, S^z_iS^z_j)\,, \label{Hlatt}
\ea
where $J'$ is the exchange coupling between the spins and ${\vec{S}}_i=(S^x_i,S^y_i,S^z_i)$ is the spin
operator at site $i$. This Hamiltonian describes a number of ferro/antiferromagnets according to $J<0$ or $J>0$, respectively. The parameter $\lambda$ accounts for the
anisotropy: for $\lambda>0$ spins tend to
align along the $z$-axis (easy-axis regime); for $\lambda=0$ we
have the isotropic case; while for $-1<\lambda<0$ one gets the
easy-plane regime. Finally, $\lambda=-1$ yields the so-called
$XY$ model (or the Planar Rotator Model, PRM, if we focus only on 2-component spin, imposing $S_z=0$).\\

In the continuum approach of spatial and spin variables, valid at sufficiently
large wavelength and low temperature, the time-independent model above may be
written as (for that, we have considered a square lattice; also $J\equiv{J'/2}$):

\ba
&H_1&=J\int\int\,\sum_{i,j=1}^{2}\;\sum_{a,b=1}^{3}\, g^{ij}h_{ab}\
(1+\delta_{a3}\,\lambda)\left( \frac{\partial S^a}{\partial\eta_i} \right)\
\left( \frac{\partial S^b}{\partial \eta_j} \right)\sqrt{|g|} d\eta_1 d\eta_2
=\nn\\
& & =J\int_\Omega\int\, (1+\delta_{a3}\,\lambda)(\vec{D}\,S^a)^2 d\Omega\,
\label{H1}
\ea
where $\Omega$ is the surface with curvilinear coordinates $\eta_1$ and $\eta_2$, so that $d\Omega=\sqrt{|g|}d\eta_1 d\eta_2$, $\delta_{ab}$ is the Kronecker symbol, $\vec{D}$ is the
covariant derivative, $\sqrt{|g|}=\sqrt{|det[g_{ij}]|}$, and $g_{ij}$ and
$h_{ab}$ are the elements of the surface and of the spin space metrics, respectively, while $g^{ij}g_{jk}=\delta^i_k$. Now,
$\vec{S}=(S_x,S_y,S_z)\equiv(\sin(\Theta)\cos(\Phi);\, \sin(\Theta)\sin(\Phi);\,
\cos(\Theta))$ is the classical spin vector field valued on a unity sphere (internal space), so that $\Theta=\Theta(\eta_1,\eta_2)$ and
$\Phi=\Phi(\eta_1,\eta_2)$. The
Hamiltonian above may be also viewed as an anisotropic non-linear $\sigma$
model (NL$\sigma$M), lying on an arbitrary two-dimensional geometry, so that
our considerations could have some relevance to other branches like
Theoretical High Energy Physics and Cosmology.\\

Here, our interest is to study the model above on the spherical geometry (the
simplest positively curved surface), so that we shall be dealing with a magnetic
shell approximated by a continuum distribution of spin sites. Thus, both the
internal (spin) and the physical spaces are identical: the sphere\footnote{Later, whenever we shall be dealing with the vortex-like solutions associated to the PRM, we shall have the internal spin space given by a circle, $S^2_x+S^2_y=S^2=1$.}. We have already parametrised the spin space by $\Theta$ and
$\Phi$. We also choose $\vartheta$ and $\varphi$ as the polar and azimuthal
angles of the magnetic sphere. Then, from the relation $x^2+y^2+z^2=R^2$ we
have $x=R\sin(\vartheta)\cos(\varphi)$, $y=R\sin(\vartheta)\sin(\varphi)$ and
$z=R\cos(\vartheta)$, so that $\Theta=\Theta(\vartheta,\varphi)$ and
$\Phi=\Phi(\vartheta,\varphi)$. Once $(R={\rm constant}, \vartheta,\varphi)$
defines an orthogonal coordinate system, then $h_{ab}$ and $g_{ij}$ may be
written as diagonal matrices; in addition, since $\vec{S}=(S_x,S_y,S_z)$ we get
$h_{ab}=\delta_{ab}$ while the metric elements
for the physical sphere reads:
$g_{\vartheta\vartheta}=g_{\varphi\varphi}=R^2$, $g_{\varphi\varphi}=R^2\sin^2(\vartheta)$, and $g_{\vartheta\varphi}=g_{\varphi\vartheta}=0$. Thus, the Hamiltonian (\ref{H1})
may be written as:

\ba
&H_{\rm sphere}&= J\int \,d\varphi \int \, d\vartheta
\left\{  \sqrt{\frac{g_{\varphi\varphi}}{g_{\vartheta\vartheta}}} \left[(1+\lambda\sin^2\Theta)\left(\partial_\vartheta
\Theta\right)^2 +\sin^2\Theta\left(\partial_\vartheta \Phi\right)^2\right]\
+\right.\nn
\\
& &+\left.
\sqrt{\frac{g_{\vartheta\vartheta}}{g_{\varphi\varphi}}} \left[(1+\lambda\sin^{ 2}\Theta)\left(\partial_\varphi \Theta\right)^2 +\sin^{ 2}\Theta \left(\partial_\varphi \Phi\right)^2\right] \right\} \nn 
\\
&=& J\int^{2\pi}_0 \,d\varphi \int^\pi_0 \, \sin(\vartheta)d\vartheta
\left\{\left[ (1+\lambda\sin^2\Theta)\left(\partial_\vartheta \Theta\right)^2 +\sin^{ 2}\Theta\left(\partial_\vartheta \Phi\right)^2\right]\right.\,+\nn
\\
& & +\left.\frac{1}{\sin^{ 2}(\vartheta)}\left[  (1+\lambda\sin^2\Theta)\left(\partial_\varphi\Theta\right)^2+ \sin^{ 2}\Theta \left( \partial_\varphi\Phi\right)^2\right]\right\}
\label{Hsphere}\,,
\ea
where $\partial_{{\eta}_i}$ accounts for $\partial/\partial\eta_i$. From
this Hamiltonian there follow the (static) equations describing the spatial distribution and dynamics of the spin fields, like below:
\ba
&\hspace{-.5cm}\left(1+\lambda\sin^2\Theta\right) \left[\partial_\vartheta( \sin\vartheta\,
\partial_\vartheta \Theta)+ \frac{\partial^2_{\varphi}\Theta}{\sin\vartheta} \right]=&\hskip -.5cm { -} \lambda\sin\Theta\,\cos\Theta
\left[\sin\vartheta (\partial_\vartheta \Theta)^2
+\frac{(\partial_\varphi\Theta)^2}{\sin\vartheta} \right]+\nn\\& & \hskip
-0.5cm+\sin\Theta\,\cos\Theta \left[\sin\vartheta (\partial_\vartheta \Phi)^2
+\frac{(\partial_\varphi\Phi)^2}{\sin{ \vartheta}} \right]\,,\label{eq1}\\ \nn\\
& \partial_\vartheta (\sin\vartheta \, \sin^2\Theta \partial_\vartheta\Phi)
+\partial_\varphi\left(\frac{1}{\sin\vartheta} \sin^2\Theta \partial_\varphi
\Phi\right)=&0\,.\label{eq2}
\ea

As expected, the general anisotropic regime of the Heisenberg model is describe
by highly non-linear differential equations. Suitable non-trivial solutions
can be obtained provided some conditions are imposed. However, before
proceeding further with such an analysis we should note that the expressions
(\ref{Hsphere}-\ref{eq2}) resemble those counterparts for the planar (zero
curvature) and for the pseudospherical (constant negative curvature) surfaces. Indeed,
whenever $R\sin\vartheta$ is identified with $r$ or $R\tau$, while $\varphi$
keeps it role as the azimuth-like angle, these expressions recover their
planar or pseudospherical analogues\cite{pseudonosso} (here, $r=|\vec{r}|$ is
the usual radial distance while $R\tau$ accounts for the distance measured
along a pseudospherical geodesic, say, a hyperbole). Such a similarity is tied
to the fact that all of these three geometries present constant Gaussian curvature
everywhere and the measurement of their respective azimuthal angles coincides.\\

To our knowledge, the isotropic Heisenberg model on the sphere was previously considered in the work
of Ref.\cite{soliton-cylinder-sphere}. There, the authors were basically interested in
cylindrically symmetric solitonic solutions, so that $\lambda=0$,
$\Theta=\Theta(\vartheta)$ and $\Phi=\Phi(\varphi)=\varphi +{\rm constant}$. In this case the Hamiltonian (\ref{Hsphere}) is simplified to:

\ba 
H_{\rm soliton}=2\pi J\int^\pi_0\left[\left(\frac{d\Theta}{d\vartheta}
\right)^2 +\frac{\sin^2\Theta}{\sin^2\vartheta}\right]\, \sin\vartheta \,
d\vartheta\,,\label{Hsoliton}
\ea
while eqs. (\ref{eq1})-(\ref{eq2}) are reduced to the unique one:
\ba 
\sin\vartheta\frac{d^2\Theta}{d\vartheta^2} +\cos\vartheta\frac{d\Theta}{d\vartheta}= \frac{\sin\Theta\,\cos\Theta}{\sin\vartheta}\,,\label{eqsoliton}
\ea
whose simplest non-trivial solutions are the ``hedgehog'' given by
$\Theta_S=\pm\vartheta$ and $\Theta_{AS}=\pi\pm\vartheta$, which have unity
topological charge ($|Q|=1$) and present finite energy equal to $8\pi{J}$ (we
refer the reader to Ref.\cite{soliton-cylinder-sphere} for further details).\\  

\section{Vortex-like solutions} 

Vortices are currently observed
in ferromagnetic nanodisks exhibiting an in-plane structure everywhere except
at its core where the spins develop out-of-plane component (see Ref.\cite{nanodot} and related references cited therein). Whenever modelling the vortex as a continuum of spins
we intend to describe only the outer region of the core, once inside the analytical
treatment is expected to give only an estimative of its energy, shedding no
light about its real structure and spins arrangement which require
numeric/simulation techniques. Here, we are interested in vortex-like excitations and how their properties are
affected by the spherical geometry. We also wish to mainly consider the so-called `in-plane' 
solutions associated to the Planar Rotator Model, whose continuum Hamiltonian may be obtained from (\ref{Hsphere}) with $\lambda=-1$ and $\Theta=\pi/2$, and reads (once we are dealing with static solutions, our results apply equally well to the XY model):
\ba
H_{\rm PRM}= J\int\int \left[ (\partial_\vartheta \Phi)^2 +\frac{1}{\sin^2\vartheta} (\partial_\varphi \Phi)^2 \right]\
\sin(\vartheta)\,d\vartheta\, d\varphi\,, \label{Hinplane}
\ea 
which yields the Laplacian equation (for the region outside the cores):
\ba
\nabla^2\Phi(\vartheta,\varphi)= \partial_\vartheta(\sin\vartheta { \partial_{ \vartheta}}\Phi)
+\frac{\partial^2_\varphi\Phi}{\sin\vartheta}=0\,. \label{eqinplane}
\ea
If $\Phi=\Phi(\vartheta,\varphi)$ then eq. above, augmented by the usual radial radial term, is solved by the well-known
spherical harmonics which present no cores (see Section 5 for further
details). The simplest core solution is obtained if we demand that $\Phi$ has
cylindrical symmetry, say, $\Phi=\Phi(\varphi)$. In this case, eq.
(\ref{eqinplane}) is easily solved by:
\ba
\Phi(\varphi)=Q\varphi +\varphi_0 \,, \label{vortexsolution}
\ea
where $Q$ is the charge of the vortex (more precisely, of one of its cores, like below) while $\varphi_0$ is a constant accounting for its global profile, giving no contribution to its energy.
The charge (vorticity) is formally defined in the continuum limit as:
\ba
Q=\frac{1}{2\pi}\oint_C (\vec{\nabla}\Phi)\cdot\,d\vec{l}\,, \label{vorticity}
\ea
where the integration is evaluated along a closed path, $C$, around the
core(s). Taking the solution above to the Hamiltonian (\ref{Hinplane}) we
obtain its energy:
\ba
E_{\rm v}=2\pi\,JQ^2\ln\left(\frac{\tan\left(\frac{\pi -\vartheta_a}{2}\right)}{\tan(\vartheta_a/2)} \right)=4\pi{J} Q^2 \ln(\cot(\vartheta_a/2))\,,\label{Evsphere}
\ea
where $\vartheta_a=a/R$ is a cutoff introduced to prevent spurious divergences
at the cores, brought about by the continuum approach. There, $a$ and $R$ are
the core and sphere radii, respectively.\\

\begin{figure}[!h]
\centering \hskip 1cm 
\fbox{\includegraphics[width=6cm]{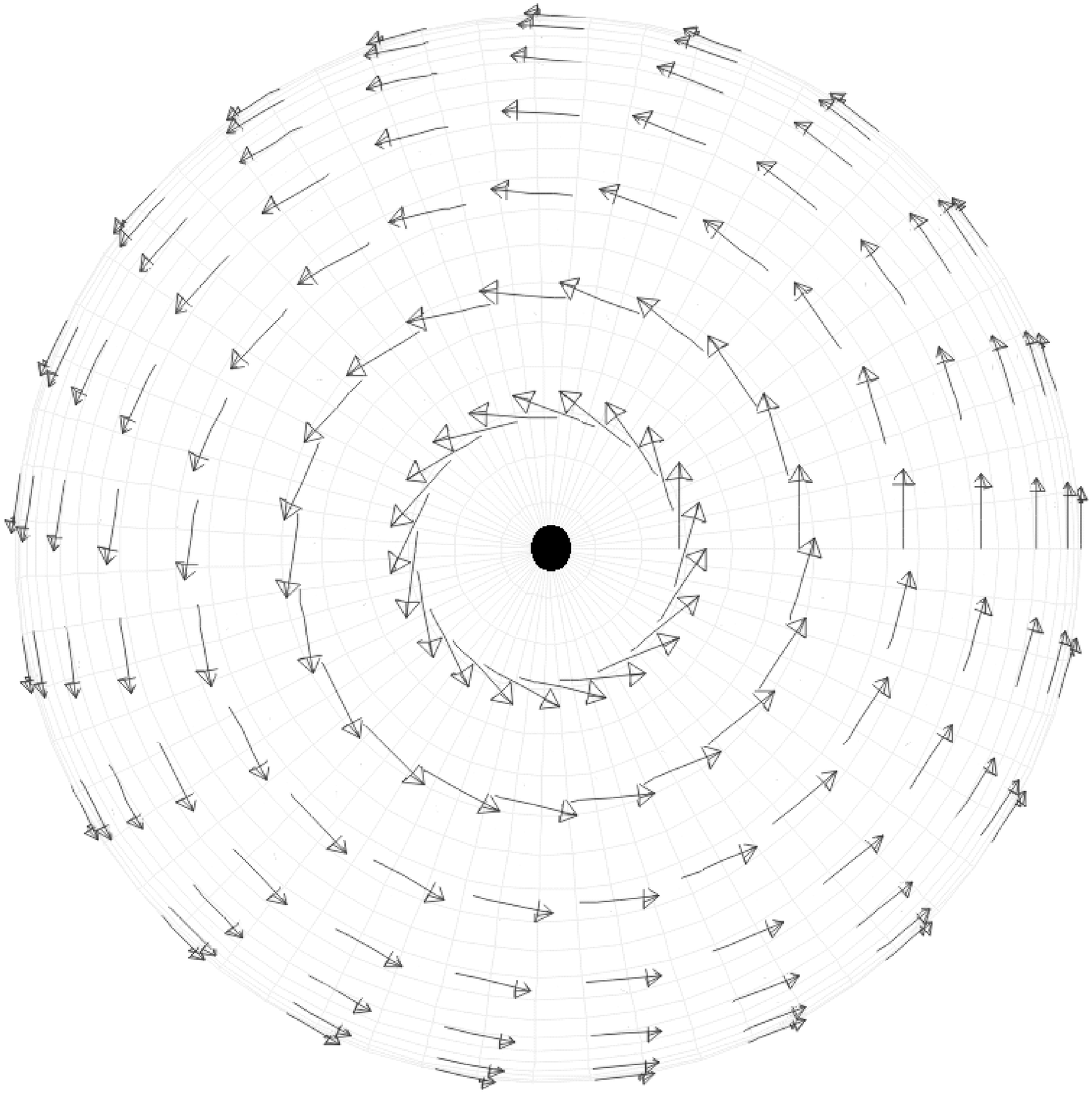} \hskip 1cm  \includegraphics[width=6cm]{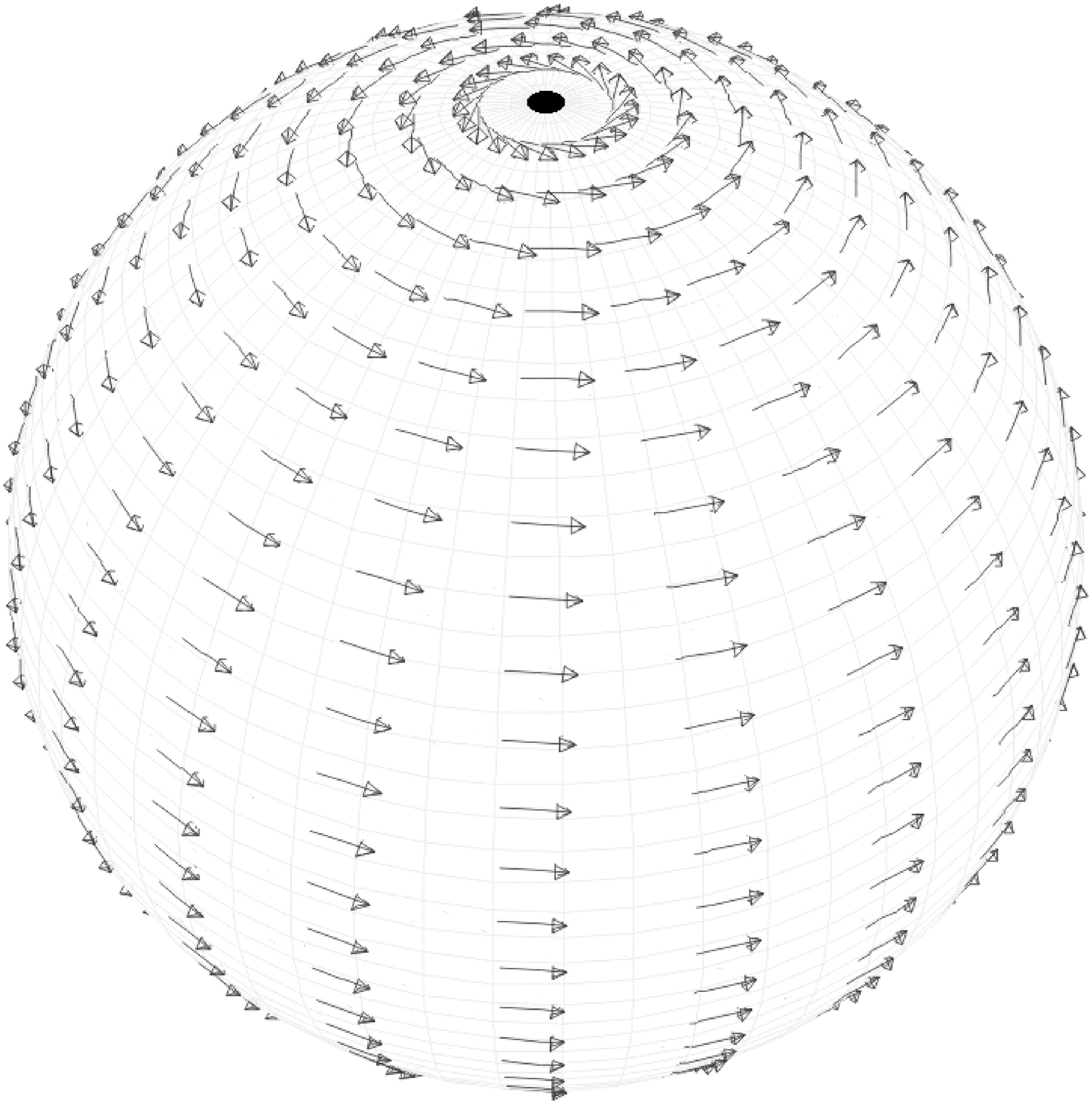}}
\caption{{\protect\small Top (left) and global (right) views of a vortex with $Q=+1$ on the sphere, showing the core at the north pole (where spins are expected to be perpendicular to the surface, for minimising energy). The arrows represent the spin field on this support, $\nabla\Phi$.}} \label{fig-vortex}
\end{figure}

Note that this vortex present two cores centred at antipodal points on the
sphere, say at $\vartheta=0$ and $\vartheta=\pi$. The appearance of the second
core is not so surprising and may be understood, for instance, if we recall that the sphere
may be stereographycally mapped to the (infinite) plane, and vice-versa.
Indeed, on the plane the vortex energy reads (in our units) $E_{\rm
v-plane}=2\pi\,JQ^2(\ln L -\ln a)$, whose terms are the large and small
cutoff, respectively. By wrapping the infinite plane to a sphere the vortex
core centred at the origin is now taken to $\vartheta=\pi$ (south pole), while all those spins located at the infinite, $L\to\infty$, are now
identified with a unique one at $\vartheta=0$ (north pole). So, the border of
the vortex at $L\to\infty$ is mapped to an extra core on the sphere (a schematic view of a vortex on the sphere is present in Fig. \ref{fig-vortex}). At the cores the spins are expected to be directed perpendicularly to the surface for minimising energy cost. Then, a more realistic picture of the vortex is that $\Theta\to 0$ as $\vartheta\to 0,\pi$ while $\Theta\to\pi/2$ outside the cores, like we have obtained here (everywhere, $\Phi=Q\varphi+\varphi_0$, as above). In this approach, the cores energy may be analytically estimated by restricting the integration in eq. (\ref{Hsoliton}) to the cores regions, say, from $\vartheta=0$ to $\vartheta_a=a/R$ and from $\pi-\vartheta_a$ to $\pi$, in such a way that the spins roughly behave as a solitonic field there ($\vartheta_a=a/R$ is the angular size of each core, like below).\\

Once $\vartheta_a$ is very small, then eq. (\ref{Evsphere}) may be
approximated to ($\cot(x)=x^{-1} +{\cal O}(x^{-2})$):
\ba
E_{\rm v}\approx 4\pi\,JQ^2 \ln (2R/a)\,,\label{Evapprox}
\ea 
so that the whole vortex may thought
as the sum of two single-core vortices, each of charge $Q$, spreading throughout a
hemisphere of radius $R$ ($R>>a$). Note also that the energy of one single-core vortex
(vortex on one hemisphere) equals that presented by its
planar counterpart in a disk of radius $2R$, so that the energy densities are the same in both surfaces. Therefore, the vortex energy on a disk is lower than on a hemisphere with the same radius. Then, if we insert a hemisphere into a plane we may expect that the former will `repeal' the vortex from its surface (see next section for details). Note, however, that such an approximation must be considered with care since the divergence of the vortex energy is not related to the sphere radius, it rather lies possibly at the cores, as correctly stated by eq. (\ref{Evsphere}).\\

\begin{figure}[!h]
\centering \hskip 1cm 

{\includegraphics[width=9cm,angle=-90]{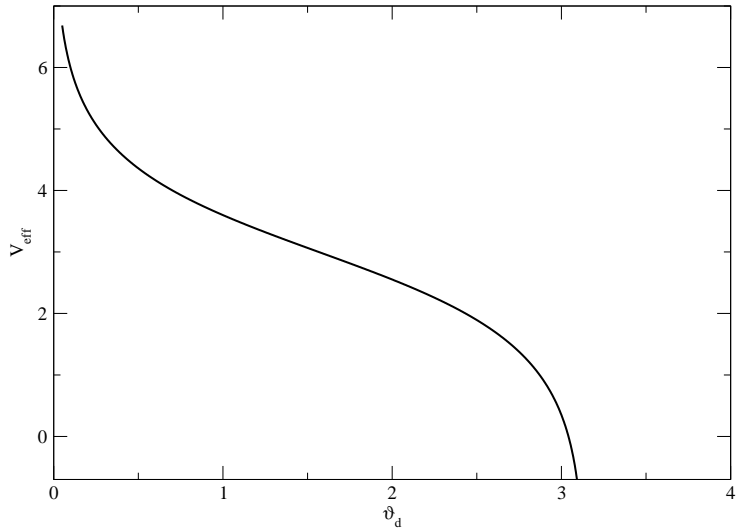}}
\caption{{\protect\small The effective potential, $V_{\rm eff}/2\pi{J}$, between the two cores with charges $Q=1$ on the sphere. Since $V_{\rm eff}$ is minimised as $\vartheta_d\to\pi$, then the cores are located at antipodal points.}} \label{figVeff}
\end{figure}

By now we have pictured our vortex as a solution with two cores, centred at
$\vartheta=0,\pi$, each of them carrying a vorticity $Q$. Then, we could be
tempted of thinking on the vortex as an excitation with net charge $2Q$.
However, the situation is quite different: as a whole the vortex has a
vanishing vorticity. Indeed, if we take two loops, each of them surrounding
each of the cores, we would measure $Q_1=Q_2=Q$. Nevertheless, the net charge
is correctly evaluated by taking one loop containing the two cores. In this
case, while one core is surrounded, say, in the $+\hat{\varphi}$ direction the
another is circulated in the opposite sense, so that the total vorticity is
$+Q + (-Q)\equiv0$. This fact is guaranteed by Gauss theorem applied to the sphere\footnote{This result may be generalised to any vortex-like solution \cite{Crowdy}, and possibly for other compact surfaces topologically (homeomorphically) equivalent to the sphere. In the latter case since the surfaces may be geometrically distinct, further attention should be given to peculiar regions/points, like those where curvature blows up, for instance the vertices of a cube, etc.}. This implies that magnetic vortices lying on a spherical support must only appear ``in pair'' with the cores centred at antipodal points. It should be stressed that the result above does not conflit with our previous analysis of vortices and their energies, once we were considering the charge of each isolated core. Indeed, eq. (\ref{Evsphere}) gives the energy of the vortex whose cores are separated by $s=\pi R$, which is the most energetically favourable configuration, like follows. For that, let us consider that the two cores are apart $d=\vartheta_d R$ (for simplicity, along a circle with $\varphi=\,{\rm constant}$; if they were aligned along different $\varphi$'s the case may be solved in the same lines). Its energy may be analytically obtained to be:\footnote{Instead of explicitly evaluating an integral similar to eq. (\ref{Hinplane}), we may use the fact that $\nabla\Phi$ is analytic everywhere, except at the vortices cores, around which $\Phi$ is a multivalued function. In this method, the result of eq. (\ref{Hinplane}) is much easier and elegantly obtained. For details, see Ref. \cite{CL}.}
\ba
E_{\rm d}=E_{\rm v1}+E_{\rm v2}+ 4\pi\,JQ_1Q_2 \ln\left(\frac{\tan\left(\frac{\pi-\vartheta_a}{2} \right)}{\tan\left(\vartheta_d/2\right)}\right)\,,
\ea
where $E_{\rm v1}$ and $E_{\rm v2}$ are the energies of the isolated cores, so that $Q_1=Q_2$, while the remaining term accounts for the effective potential, $V_{\rm eff}$, between them. For preventing spurious divergences we take the angular separation between the isolated cores as $\vartheta_d\in[\vartheta_a,\pi-\vartheta_a]$. Since $V_{\rm eff}$ monotonically decreases with increasing $\vartheta_d$ in the range above, then the lowest energy configuration demands the cores to be centred at antipodal points, say, separated by $\vartheta=\pi$ (see Fig. \ref{figVeff}). Multi-vortex solutions may be considered in the same way. In this case, we can have cores with opposite charges so that these do attract each other, making the scenario more interesting. In the simplest case, we have two pairs ($Q_+=+1$ and $Q_-=-1$) very close, say, around the north and  south poles. An interesting issue to be studied is the thermodynamics involved in this dissociation of pairs and how this could change the spin-spin correlation function in this geometry.\\ 

\section{Slow vortex dynamics on a surface with interfaces}

\begin{figure}[!h]
\centering \hskip 1cm 
{\includegraphics[width=10cm]{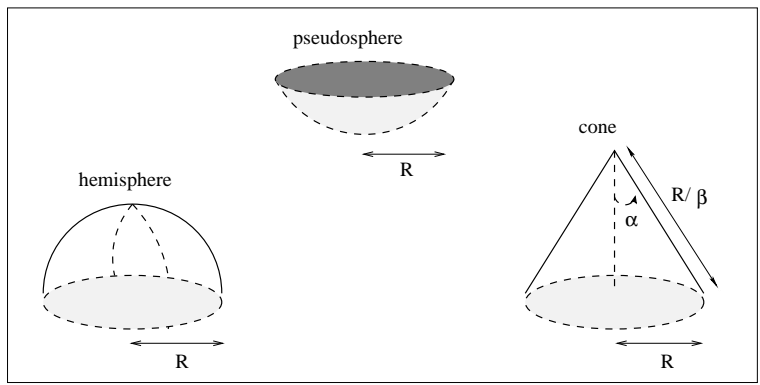}}
\caption{{\protect\small A schematic view of the segments in the plane.}} \label{figinterfaces}
\end{figure}

For the sake of completeness, we would like to consider the problem of how a hemisphere may affects the dynamics of a magnetic vortex. For comparison, we also consider the effect of a cups-like defect of conical shape and of a pseudospherical segment (the situation is schematically shown in Fig.\ref{figinterfaces}). For that, we consider a large magnetic plane in which these geometries, with base radii $R$, are introduced. For simplifying the analysis we suppose that: i) the interfaces between the plane and each segment is smooth enough, so that their effects are not appreciable; ii) the inserted segments are sufficiently apart, exerting no influence each other, and; iii) the vortex motion is slow (adiabatic) and induces no changes in its cylindrically symmetric and planar profiles, like deformation near interfaces, development of out-of-plane component, etc. At some extent, our analysis is an attempt to carry out the work of Ref.\cite{SaxenaPLA}, where solitonic-like excitations in an line with curved sectors was considered, to two dimensions.\\

\begin{figure}[!h]
\centering \hskip 1cm 
{\includegraphics[width=8cm,angle=-90]{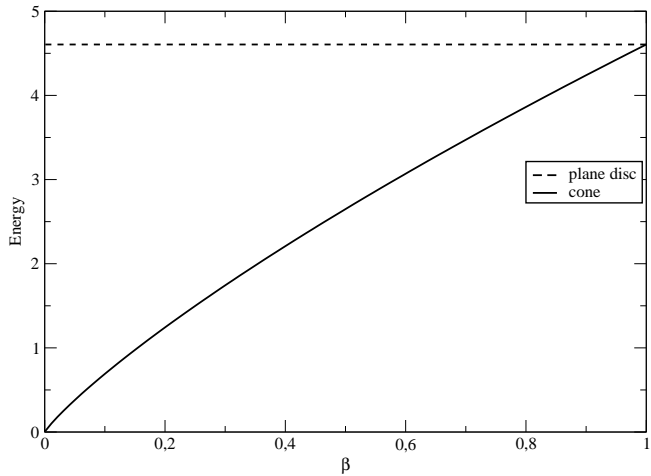}}
\caption{{\protect\small The energy of the vortex (divided by $2\pi{J}Q$) as function of $\beta$. Note that, for every $\beta<1$ its value on the cone (solid curve) is lower than in the flat disc plane (dashed line).}} \label{energy-plane-cone}
\end{figure}

With these assumptions, only the vortex energy outside the core must be computed for our purposes. First, suppose that the vortex core is placed at the conical apex. In this case, its total energy reads (plane' in the subscript stands for the plane minus the disc area, $\pi R^2$):
\ba
E_1= \int_{\rm cone}(\nabla\Phi)^2 d\Omega +\int_{\rm plane'}(\nabla\Phi)^2 dxdy= 2\pi{J}Q\left[\beta\ln(R/{\beta}a)+\ln{L/R}\right]\,,
\ea
which is less than $E_{\rm plane}=2\pi{J}Q^2\ln(L/a)$, whenever $\beta<1$ (see Fig. \ref{energy-plane-cone}). Above, $R/\beta$ is the size of the cone and $\beta=\sin\alpha$ with $\alpha$ being the cone half-aperture angle. Therefore, it is energetically favourable to find the vortex core pinned at the apex (this attraction was previously studied in the work of Ref.\cite{vortex-cone}, where further details upon the conical geometry and parameters may be obtained).\\

Now, let us see what is the effect of the hemisphere. In this case, the vortex energy reads:
\ba
E_2= \int_{\rm hemisphere}(\nabla\Phi)^2 d\Omega +\int_{\rm plane'}(\nabla\Phi)^2 dxdy= 2\pi{J}Q\left[\ln(2R/a)+\ln(L/R)\right]\,.
\ea
Now, $E_2>E_{\rm plane}$ implying that the vortex is repealed from the hemisphere.\\

Let us consider now a pseudospherical segment, of base radius $R$, inserted in the plane. In this case, the distance measured along a hyperbole from the origin to the interface with the plane is given by $s_R=r'\tau_R/2$ where $r'$ is a parameter introduced for ensuring length dimension (Poincar\'e disc radius) while $\tau_R=\arcsin(2R/r')$. Therefore, the energy of the vortex if it were centred at the pseudospherical origin is given by:\footnote{For determining $\tau_R$ we take the projection of the pseudosphere segment on the $xy$-plane. This is given by $x^2+y^2=R^2$ where $x=r'\cos(\varphi)\sinh(\tau)$ and $y=r'\sin(\varphi)\sinh(\tau)$ while $\varphi$ and $\tau$ are the pseudospherical coordinates. In general, distances measured along the geodesics (hyperbola) are given by $s=r'\tau/2$. For further details upon this geometry see Ref.\cite{pseudonosso}.}
\ba
E_3=\int_{\rm pseudosphere}(\nabla\Phi)^2 d\Omega +\int_{\rm plane'}(\nabla\Phi)^2 dxdy=2\pi{J}Q\left[\ln\left(\frac{\tanh[\arcsin(R/r')]}{\tanh(a/r')}\right)+\ln(L/R)\right]\,,
\ea

\begin{figure}[!h]
\centering \hskip 1cm 
{\includegraphics[width=8cm,angle=-90]{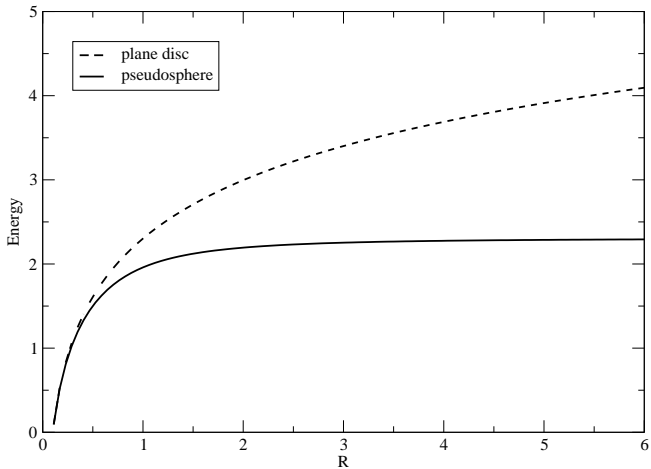}}
\caption{{\protect\small The energy of the vortex (divided by $2\pi{J}Q$) as function of $R$. Note that its value on the pseudosphere (solid line) is lower than in the flat disc (dashed curve), namely for $R$ large enough.}} \label{energy-plane-pseudo}
\end{figure}

\noindent where $a$ is the actual vortex core size on this geometry. Similarly to what happens in the conical case, $E_3<E_{\rm plane}$, so that the vortex tends to be attracted and centred at the pseudospherical origin (see Fig. \ref{energy-plane-pseudo}). In conclusion, vortices in a magnetic plane, with hemispherical, pseudospherical and conical deformations sufficiently apart, are expected to be centred at the apex or at the saddle point of the pseudosphere (whether they actually choose the apex or the pseudosphere depends further on $\beta$ and $R$ parameters; such an analysis may be easily carried out). Although obtained from a very simple analysis it is worthy to mention that our results qualitatively agree which those observed with vortex lines in helium superfluids\cite{Voll-Zieve}.\\

\section{Coreless solutions: spherical harmonics spin fields}

In Section 3 we have studied the simplest cylindrically symmetric solutions of the equations of motion. This led us to realize that on a sphere a vortex presents two cores with same charge at antipodal points while the net vorticity identically vanishes. Now, if we relax the demand for such a symmetry and search for $\Phi=\Phi(\vartheta,\varphi)$ then we obtain the well-known spherical harmonics, say (for that we must start off from the complete 3D Laplacian, with radial dependence, and after all we take $r=R={\rm constant}$):
\ba
\Phi(\vartheta,\varphi)=Y^m_n(\vartheta,\varphi)=N\, P^m_n(\vartheta)\, e^{im\varphi}\,,\label{harmonicos}
\ea
where $N\equiv(-1)^m \sqrt{\frac{2n+1}{4\pi}\frac{(n+m)!}{(n-m)!}}$ and $P^m_n$ are the associated Legendre polynomials, defined in terms of the Legendre polynomials, $P_n(x)$, by $P^m_n(x)=(1-x^2)^{m/2}\, \frac{d^m}{dx^m} P_n(x)$, with $x=\cos(\vartheta)$. In addition, $-n\le m \le+n$ and to guarantee single-valuedness of $Y^m_n$ in $\varphi$, $m$ is integer.\\

\begin{figure}[!h]
\centering \hskip 1cm 
\fbox{\includegraphics[width=6cm,height=6cm]{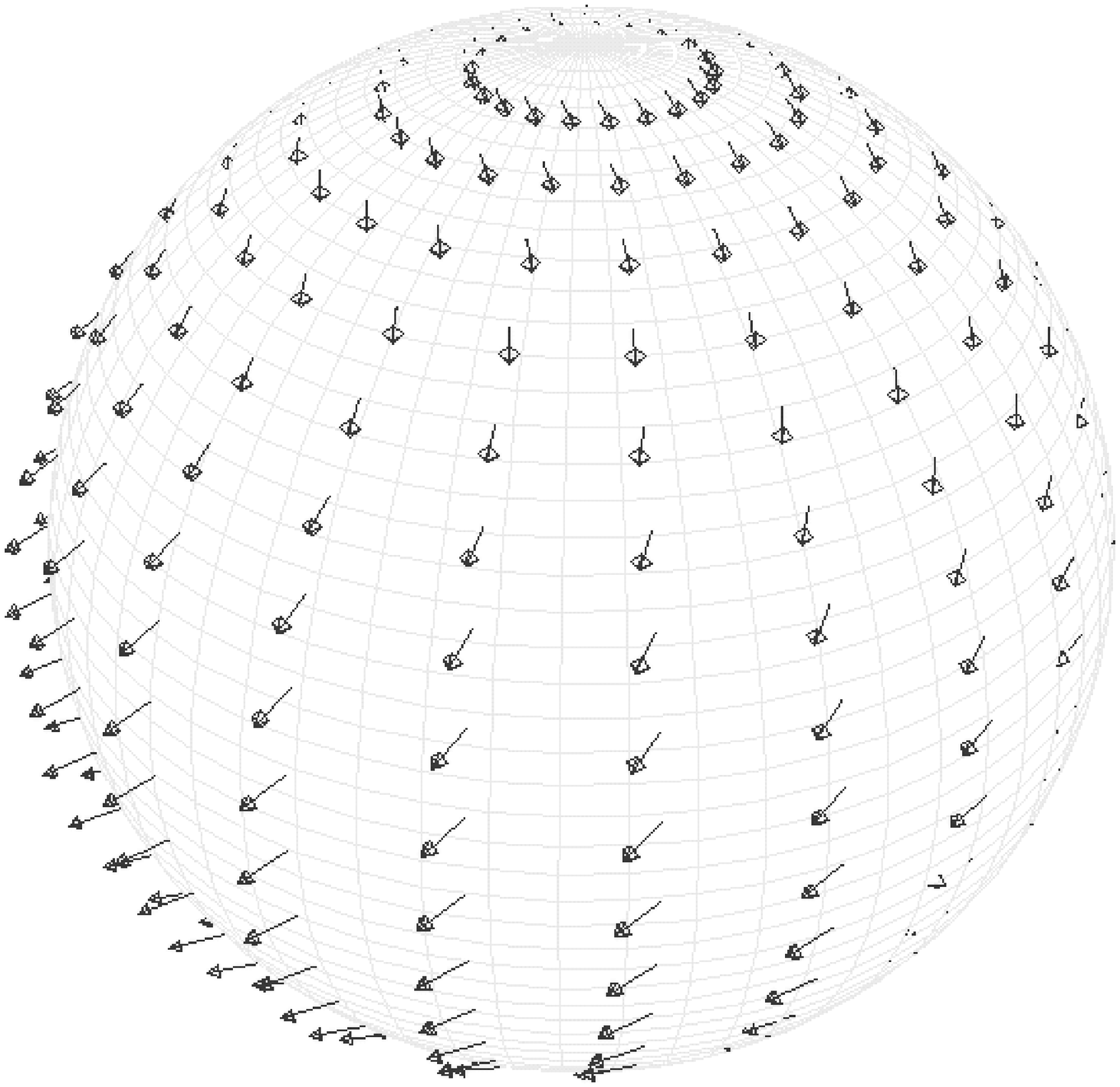}} \hskip 1cm  \fbox{\includegraphics[width=6cm,height=6cm]{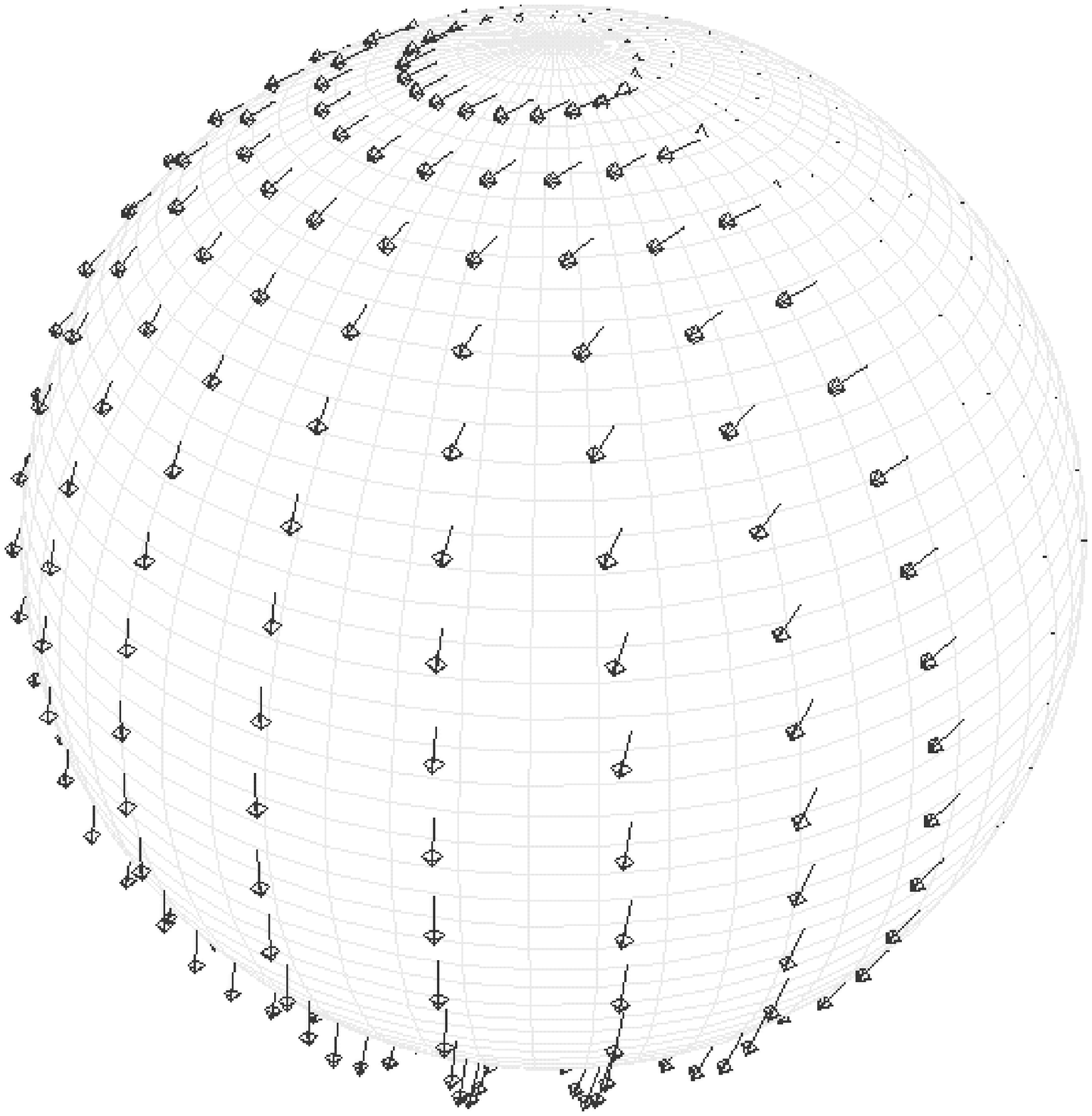}}
\caption{{\protect\small Schematic views of the two simplest spherical harmonics coreless spin field solutions. On the left we have $Y^0_1$ while on the right $Y^1_1$ is showed.}} \label{figharmonic}
\end{figure}

Clearly, $\nabla\times(\nabla Y^m_n)\equiv0$ everywhere so that no singular cores are now present. Some plots of the vector field $\nabla Y^m_n$ are displayed in Fig. \ref{figharmonic}. Although $m$ is important for its profile it is immaterial for the associated energy, like below:
\ba
E_n=J\int(\nabla\Phi)^2d\varphi\, \sin(\vartheta)d\vartheta= 2\pi{J}N^2 \int^\pi_0 \sin(\vartheta) d\vartheta \left[\left(\frac{d}{d\vartheta}P^m_n \right)^2 +\frac{m^2}{\sin^2(\vartheta)}(P^m_n)^2\right]\,,\nn
\ea
where $P^m_n=P^m_n (\cos(\vartheta))$. Despite of its form this integral is easily solved for yielding:\footnote{We have used the identity\cite{Arfken}:$$\int^\pi_0 \sin(\vartheta) d\vartheta \left[\left(\frac{d}{d\vartheta}P^m_n (\cos(\vartheta))\right) \left(\frac{d}{d\vartheta}P^m_l (\cos(\vartheta))\right) +\frac{m^2}{\sin^2(\vartheta)}P^m_n(\cos(\vartheta))P^m_l(\cos(\vartheta))\right]= \frac{2n(n+1)}{2n+1}\frac{(n+m)!}{(n-m)!}\delta_{nl}\,.$$}
\ba
E_n=n(n+1)J \,.\label{energyharmonics} 
\ea
Whether such regular structures appear and are stable on a magnetic sphere remains to be further investigated, for instance, by simulation and experiments. At principle, a point favouring their formation seems to be their finite energy, similar to what happens to topologically stable soliton-like excitations. On the other hand, energetics tends to prefer the more fundamental solutions, say, with small $n$ values.\\

\section*{Conclusions and Prospects}

The simplest magnetic vortex-like solution of the Planar Rotator Model, on a spherical surface, present two singular cores, with the same charge, which repeal each other so that they are located at antipodal points, for instance, at the north and south poles. In addition, its associated (finite) energy does not depend upon the sphere radius so that it does not blow up as the thermodynamic limit is taken, $R\to\infty$. These points should be contrasted with the common solution displaying an unique core, like those observed in open surfaces, whose energy diverges with the system size according $\ln(L)$.\\

Furthermore, we consider a problem of slow  vortex motion on a plane with inserted segments of conical, hemispherical and pseudospherical shapes. We realise that while the excitation is attracted by the conical apex and by the centre (origin) of the pseudosphere it is repealed by the hemisphere, suggesting these surfaces as pinning and depinning geometries for magnetic vortices. Attention was also paid to everywhere regular (coreless) solutions which are the well-known spherical harmonics. Their energies were analytically computed and shown to be finite, resembling at some extent, those associated to soliton-like excitations. Whether such solutions could be formed and are stable is discussed in some details.\\

As prospects for future investigation we may quote the analysis of multi-vortex solution, mainly in connection with a possible topological phase transition associated to the vortex-antivortex dissociation in this geometry\cite{workinprogress}. Similar issue has been considered in the conical case, where the temperature transition seems to be shifted below as the cone aperture is decreased ($\beta$ is lowered) \cite{vortex-cone} and on the pseudospherical support where vortex dissociation seems to take place at any arbitrary small temperature \cite{pseudonosso}. The study of vortices in other geometries, like the torus, could also shed some light to the problem of phase transition. For instance, it has been recently verified in simulations that the two-dimensional Ising model defined on a toroidal lattice present two critical temperatures\cite{shimatoro} instead of one, like in the usual planar case. The effects of holes (impurities/defects) on topological excitations lying on the spherical surface is another interesting subject to be worked out. For instance, hollowed nanospheres with vortex-like magnetisation could be important for potential applications, namely if they were considered at the nanoscale.\\ 

\vskip 1cm
\centerline{\large\bf Acknowledgements} \vskip .7cm
The authors are grateful to A.R. Moura and S.G. Alves for computational help. L.A.S. M\'ol and N.M. Oliveira-Neto are acknowledged for fruitful discussions. Professor Gary Williams (UCLA) is also acknowledged for having drawn our attention to some important references. They also thank CNPq and FAPEMIG for the financial support.


\vskip 1cm

\thebibliography{99}

\bibitem{sphere-new-methods} D.R. Nelson, Nano Lett. {\bf 2}, 1125 (2002);\\
R.D. Kamien, Science {\bf 299}, 1671 (2003);\\
A.R. Bausch, M.J. Bowick, A. Cacciuto, A.D. Dinsmore, M.F. Hsu, D.R. Nelson,
M.G. Nikolaides, A. Travesset, and D.A. Weitz, Science {\bf 299}, 1716 (2003).

\bibitem{Prodan} E. Prodan, C. Radloff, N.J. Halas, and P. Nordlander, Science
{\bf 302}, 419 (2003);\\
E. Prodan, P. Nordlander, and N.J. Halas, Nano Lett. {\bf 3}, 1411 (2003);\\
E. Prodan and P. Nordlander, Nano Lett. {\bf 3}, 543 (2003).

\bibitem{sphere-superc-vortex} E.T. Filby, A.A. Zhukov, P.A.J. de Groot, M.A. Ghanem, P.N. Bartlett, and V.V. Metlushko, App. Phys. Lett. {\bf 89}, 092503 (2006).

\bibitem{vortex-chain-sphere} P. Barpanda, T. Kasama, R.E. Durnin-Borkowski,
M.R. Scheinfein, A.S. Arrott, J. App. Phys. {\bf 99}, 08G103 (2006).

\bibitem{sphere-cr-sizes} D. Goll, A.E. Berkowitz, and H.N. Bertram, Phys. Rev. {\bf B70}, 184432 (2004).

\bibitem{Vitelli-Nelson} V. Vitelli and D.R. Nelson, Phys. Rev. {\bf E74}, 021711 (2006). See also related references cited therein.

\bibitem{BKT}V.L. Berezinskii, Sov. Phys. JETP {\bf 32} 493 (1970); {\em ibid}
{\bf 34}, 610 (1972);\\
J.M. Kosterlitz and D.J. Thouless, J. Phys. {\bf C6}, 1181 (1973).

\bibitem{nanodot} M. Rahm, J. Stahl, W. Wegscheider, and D. Weiss, App. Phys.
Lett. {\bf 85}, 1553 (2004);\\
M. Rahm, J. Stahl, and D. Weiss, App. Phys. Lett. {\bf 87}, 182107 (2005);\\
A.R. Pereira, A.R. Moura, W.A. Moura-Melo, D.F. Carneiro, S.A. Leonel, and P.Z.
Coura,  J. Appl. Phys. {\bf 101}, 034310 (2007).

\bibitem{BP}A.A. Belavin and A.M. Polyakov, JETP Lett. {\bf 22}, 245 (1975),
see also R. Shankar, J. Phys. (Paris) {\bf 38}, 1405 (1977).

\bibitem{skyrmion} T.H.R. Skyrme, Nucl. Phys. {\bf31}, 556 (1962).

\bibitem{tHP} G. 't Hooft, Nucl. Phys. {\bf B79}, 276 (1974);\\
A.M. Polyakov, JETP Lett. {\bf 20}, 194 (1974); Soviet Physics JETP {\bf 41}, 988 (1976).

\bibitem{soliton-cylinder} S. Villain-Guillot, R. Dandoloff, and A. Saxena,
Phys. Lett. {\bf A188}, 343 (1994);\\
R. Dandoloff, S. Villain-Guillot, A. Saxena, and A.R. Bishop, Phys. Rev. Lett.
{\bf 74}, 813 (1995);\\
R. Dandoloff and A. Saxena, Eur. Phys. J. {\bf B29}, 265 (2002);\\
E.A. Silva and A.R. Pereira, Phys. Status Solidi {\bf B213}, 481 (1999); Solid
State Commun. {\bf 113}, 699 (2000).

\bibitem{soliton-cylinder-sphere}S.Villain-Guillot,  R. Dandoloff, A. Saxena,
and A.R. Bishop, Phys. Rev. {\bf B52}, 6712 (1995).

\bibitem{soliton-cone} A.R. Pereira, J. Magn. Mag. Mat. {\bf 285}, 60 (2005);\\
W.A. Freitas, W.A. Moura-Melo, and A.R. Pereira, Phys. Lett. {\bf A336}, 412
(2005);\\
V.B. Bezerra, C. Romero, and S. Chervon, Int. J. Mod. Phys. {\bf D14}, 1927
(2005);\\
A. Saxena and R. Dandoloff, Phys. Rev. {\bf B66}, 104414 (2002), for the case of a truncated cone.

\bibitem{vortex-cone} W.A Moura-Melo, A.R. Pereira, L.A.S. M\'ol, and A.S.T. Pires, Phys. Lett. {\bf A360}, 472 (2007).

\bibitem{pseudonosso} L.R.A. Belo, N.M. Oliveira-Neto, W.A. Moura-Melo, A.R. Pereira, and E. Ercolessi, ``{\em Heisenberg model on a space with negative curvature: Topological spin textures on the pseudosphere}'', Phys. Lett. {\bf A} (2007) [in press], {\bf doi:10.1016/j.physleta.2007.01.044} .

\bibitem{Crowdy} D. Crowdy and M. Cloke, Phys. Fluids {\bf 15}, 22 (2003);\\
D. Crowdy, J. Fluid Mech. {\bf 498}, 381 (2004).

\bibitem{SaxenaPhysA} A. Saxena, R. Dandoloff, and T.
Lookman, Physica {\bf A261}, 13 (1998).

\bibitem{CL} P.M. Chaikin and T.C. Lubenski ``{\em Principles of Condensed
Matter Physics}'', Cambridge Univ. Press; 1st edition (1995), see mainly
Section 9.3;\\
G.A. Willians and E. Varoquaux, J. Low Temp. Phys. {\bf 113}, 405 (1998).

\bibitem{SaxenaPLA}R. Dandoloff and A. Saxena, Phys. Lett. {\bf A358}, 421 (2006).

\bibitem{Voll-Zieve} P. Voll, N. apRoberts-Warren, and R.J. Zieve, ``{\em Surface curvature and vortex stability}'', cond-mat/0601395.\\ V. Vitelli and A.M. Turner, Phys. Rev. Lett. {\bf 93}, 215301 (2004). 

\bibitem{Arfken} G.B. Arfken and H.J. Weber, ``{\em Mathematical Methods for Physicists}'', sixth edition,  Elsevier Academic Press, 2005; see Chapter 12, mainly pages 771-802.

\bibitem{workinprogress}  V. Kotsubo and G.A. Williams, Phys. Rev. {\bf B33}, 6106 (1986);\\ H. Cho and G.A. Williams, J. Low Temp. Phys. {\bf 110}, 533 (1998);\\G.A. Williams, Phys. Rev. {\bf B73}, 214531 (2006);\\ W.A. Moura-Melo, work in progress.

\bibitem{shimatoro}I. Hasegawa, Y. Sakaniwaa and H. Shima, J. Mag. Magn. Mat.  {\bf 310} (2007) 1407.
\end{document}